\documentclass[%
  twocolumn,
 showpacs,
 showkeys,
 preprintnumbers,
 amsmath,amssymb,
 aps,
  pra,
  longbibliography,
 ]{revtex4-1}

\usepackage{mathptmx}

\usepackage{amssymb,amsthm,amsmath}

\usepackage{tikz}
\usepackage[breaklinks=true,colorlinks=true,anchorcolor=blue,citecolor=blue,filecolor=blue,menucolor=blue,pagecolor=blue,urlcolor=blue,linkcolor=blue]{hyperref}
\usepackage{graphicx}
\usepackage{url}

\usepackage{xcolor}

\usepackage[american,greek]{babel}

\begin{document}
\selectlanguage{american}

\title{Locating possible sources of physical indeterminism}


\author{Karl Svozil}
\affiliation{Institute for Theoretical Physics, Vienna
    University of Technology, Wiedner Hauptstra\ss e 8-10/136, A-1040
    Vienna, Austria}
\email{svozil@tuwien.ac.at} \homepage[]{http://tph.tuwien.ac.at/~svozil}

\keywords{Ramsey theory, statistics, providence, free will, determinism, indeterminism, choice, agent, oracle, miracle}

\begin{abstract}
Some possible (re)sources of indeterminism and randomness encountered in physics are enumerated. These gaps in the physical laws, if they exist, could possibly be exploited for dualistic interfaces. We also speculate that physical laws and order could possibly emerge from primordial chaos by means resembling statistics and Ramsey theory.
\end{abstract}

\maketitle

\section{Introduction}

Occasionally physics is  confronted with the following issue:
although there is no apparent law determining the behavior of certain physical systems,
these physical systems evolve nevertheless.
This gives rise to speculations that there are (irreducible) gaps in the (known) laws of physics,
which in turn motivate claims of indeterminism or even total randomness in nature.

Presently the consensus among quantum physicists and philosophers of science appears to be this:
as long as one considers a (pure quantum) state (of knowledge) of individual particles subsumed under the name {\em quanta},
or average properties (of ``large'' groups) of quanta,
the universe appears lawful.
Alas the details of the occurrence of {\em individual} events related to such quanta cannot be causally explained and predicted.
In particular, the details of the behavior of individual quanta -- such as the exact time of their
creation, scattering and annihilation --
are believed to be irreducibly unpredictable and random.
Thus quantum indeterminism is postulated to be ontic, as opposed to the epistemic, means relative,
way of perceiving unpredictability in classical (statistical) physics~\cite{Myrvold2011237}.

For the sake of an example, suppose we are dealing with an atom in an excited state
which eventually is bound to decay, thereby emitting a photon; that is, a quantum of light:
while the half-life (which is a probabilistic quantity)
of the excited state can be determined with
arbitrary precision, the exact time of the individual spontaneous decay remains unpredictable to this day.

Already in 1926, Born \cite[p.~302]{jammer:89}
subsumed this situation as follows:
{\it ``the motion of particles conforms to the laws of probability, but the probability itself
is propagated in accordance with the law of causality.
[This means that knowledge of a state in all points in a given time determines the distribution of
the state at all later times.]''}
In that same year Born stated that \cite[p.~54]{wheeler-Zurek:83}
{\it ``I myself am inclined  to give up determinism in the world of atoms.''}

This presents a challenge for physics, philosophy, as well as religion.
Because how should one cope with questions such as indeterminacy and the apparent violation of the principle of sufficient reason?
What kind of deity and mind is consistent with Born's inclinations?
And why does the universe appear lawful if it is essentially made up of randomly occurring events?

In what follows we shall be dealing with some aspects and alternative answers to these questions.
We shall first present some nomenclature,
followed by subjective motivation to pursue metaphysical questions of existence pertinent to the topics discussed,
and present a very brief historic account on physical unknowables~\cite{svozil-07-physical_unknowables}.
Later on we shall evaluate various sources of physical indeterminism which could potentially serve as
gaps in a universe which otherwise appears lawful.

We also shortly dive into Ramsey theory which assures us that any sufficiently large structure inevitably
contains ordely substructures which can be conjectured to be ``lawful''  -- just as the Elders looked up into the skies and
``found'' animal constellations there~\cite{GS-90}.
Note also that~\cite[Theorem~6.1, p.~148]{calude:02}, {\em ``almost all real numbers, when expressed in any base,
contain every possible digit or possible string of digits''} -- even entire deterministic universes.

A {\it caveat:} when it comes to  physical (in)determinism
one should be aware of what Jaynes called
the ``Mind Projection Fallacy'' \cite{jaynes-89,jaynes-90}:
{\em ``we are all under an ego-driven temptation to project our private
thoughts out onto the real world, by supposing that the creations of one's own imagination are real
properties of Nature, or that one's own ignorance signifies some kind of indecision on the part of
Nature.´´}

A second {\it caveat:}
with regard to determinism {\it versus} indeterminism,
there will never be any formal or empirical certainty one way or another besides the assurances of the orthodoxy.
Because for embedded observers
(as well as for ``strong enough'' formal systems supporting universal computation --
in what follows the Term ``computability''
stands for Turing computability~\cite{odi:89,odi:99})
the general prediction and rule inference (induction) problems are provable unsolvable~\cite{go-67}.

\subsection{Nomenclature}

What do we mean by unpredictable, (irreducibly) indeterministic, or even random physical behavior~\cite{2014-nobit}?
Although heuristically evident this is by no means trivial, because, for the sake of boggling the mind,
suppose that time is a continuum. Suppose further
that a single physical event (such as the emission of a photon) can happen at any time.
Then with probability one it will not happen at any computable time.
This is due to the fact that, ``almost all such times should be random.''
(Formally: since the set of computable numbers is countable,
the Lebesgue measure of computable reals on the unit interval is zero.)

In what follows we mean by {\em predictable} that there is a computable function,
in particular one with computable rate of convergence,  which,
given an arbitrary precision,
yields a prediction.
That is, the resources required for making predictions with better and better accuracies could be enormous
as long as they remain finite and computable.
(This is, for instance, not true for a {\em Chaitin Omega number}~\cite{2002-glimpseofran}:
although it is computably enumerable -- that is, the limit of a computable,
increasing, converging sequence of rationales -- it is random and, in particular,
unpredictable because it lacks a computable rate of convergence.)

Since theological nomenclature hardly belongs to the standard repertoire of physicists but will be used later,
some {\it termini technici} will be mentioned upfront.
Thereby we will mainly follow Philipp Frank's (informal) definitions of {\em gaps} and {\em miracles}~\cite{frank,franke},
as well as Robert Russell's notions allowing him
to formulate the doctrine of {\em non-interventionist objective divine action (NIODA)}~\cite[Part~II,Chapter~4]{Russel-nioda-1}.

In the theological context,  {\it creatio ex nihilo} often refers to the `initial boot up of the universe;'
whereas {\it creatio continua} stands for the permanent intervention of the divine throughout past, present, and future.
Alas, as we will be mainly interested with physical events, we shall refer to
{\it creatio ex nihilo}, or just {\it ex nihilo,} as something coming from nothing; in particular, from no intrinsic~\cite{svozil-94} causation
(and thus rather consider the theological {\it creatio continua}; apologies for this potential confusion).
{\it Ex nihilo} denies, and is in contradiction, to the {\em principle of sufficient reason}, stating that nothing is without intrinsic causation, and {\it vice versa}.

According to Frank~\cite[Sect.~II,~12]{frank,franke}, a {\em gap} stands for the {\em incompleteness} of the laws of nature,
which allow for the occurrence of events without any unique natural (immanent, intrinsic) cause, and for the possible intervention of higher powers~\cite[Sect.~II,~9]{frank,franke}: {\em ``Under
certain circumstances they do not say what definitely has to happen
but allow for several possibilities; which of these possibilities comes
about depends on that higher power which therefore can intervene
without violating laws of nature.''}

This is different from a direct breach or `rapture' of the laws of nature~\cite[Sect.~II,~10]{frank,franke};
also referred to as {\em ontological gap} by a forced {\em intervention} in the otherwise uniformly causal connection of events~\cite[Sect.~3.C.3, Type~II]{Russel-nioda-1}.
An example for an ontological gap would be the sudden {\it ad hoc} turn of a celestial object which
would otherwise have proceeded along a trajectory governed by the laws of inertia and gravitation.

Often, the resulting correlations are subjectively and semantically experienced as {\em synchronicity},
that is, with a {\em purpose} -- the events are not causally connected but {\em ``stand to one
another in a meaningful relationship of simultaneity''}~\cite{jung1,jung1e}.
A more personal example is Jung's experience of a solid oak table suddenly split right across,
soon followed by a strong steel knife breaking in pieces for no apparent reason~\cite[pp.~111-2, 104-5]{jung-memories,jung-memories-e}.

In what follows we shall adopt Frank's conceptualization of a {\em miracle}~\cite[Sect.~II,~15]{frank,franke}
as a {\em gap} (in Frank's sense cited above) which is exploited according to a {\em plan}  or purpose;
so a `higher power' interacts {\it via} the incompleteness (lack of determinacy) of the laws of nature to pursue an intention.

Note that this notion of miracle is different from the common acceptation quoted by Voltaire,
according to which a miracle is the violation of divine and eternal laws~\cite[Sect.~330]{voltaire-dict}.
Russel refers to the latter as `miracle in the Humean sense'~\cite[Sect.~3.C.7 ]{Russel-nioda-1}: {\em ``a miracle is an event which violates the laws of nature and which contradicts science.''}

An {\em oracle} is an agent capable of a {\em decision} or an {\em emanation} (such as a random number) which cannot be produced by a universal computer.
Again, we take up Frank's conception of a gap, or of Russel's NIODA, to conceptualize physical oracles.

Finally, in what follows the term {\em transcendence} refers to an entity or agent beyond all physical laws.
(It is not used in the Kantian sense.)
In contrast, {\em immanence} refers to all operational, intrinsic physical
means available to embedded observers~\cite{toffoli:79,svozil-94} from within some universe.

\subsection{The mind-boggling fact of existence}

Our first and foremost existential problem appears to be that of {\em existence}~\cite{holt-existence}: why is there something rather than nothing?
In particular, why does the universe exist? What are we here for? What is our origin and our destiny?
(I would therefore disagree with Camus~\cite{camus-mos} that the only serious philosophical problem is whether to commit suicide.)
It might not be too speculative to suggest that the resulting mind-boggling amazement is the root of both religion and science,
which share a single goal, and thus might be seen as two sides of the same endeavor: the pursuit of truth -- how ephemeral this may appear.

Reactions to existence can turn into feelings of terror, anxiety, panic or dread;
a {\it  pavor nocturnus} (night terror) in full waking life.
These are mostly stipulated by the suspicion that there are no grounds on which to anchor possible answers;
the issue of existence rises before us like an barrier insurmountable for rational thought.
Indeed, contemplation on existence, if pursuit honestly and consequentially,
may result in madness and total destructive dissolution of the self; without any hope of return to normality or redemption.

Various strategies have been developed to cope with the individual human experience of existence.
We may group these strategies into three branches: (i) religion, (ii) philosophy (e.g. existentialism, materialism), and (iii) science.
As all human endeavors, these three branches share a common ground: all of them are narratives.

Religion is a very powerful narrative, and it is a very big grace to be able to believe.
Because all of a sudden, when viewed with belief, the Universe becomes light and bright, and full of deep meaning.

Natural sciences is a less gratifying narrative, but the resulting recommendations may have, on the average, more practical usefulness.
Alas, as philosophers of science (e.g. Lakatos, Popper \& Feyerabend) clearly have stated,
there is no absolute truth in ``scientific understanding.''
Even mathematics, such as Ramsey theory, cannot help, because it itself is a formalism.

Many so-called `philosophical' approaches such as atheism and materialism appear to be mere ideologies in disguise.
(An exception is a Camus-type existentialism.)
Indeed, these ideologies may become utterly dangerous in times of political revolution and unrest (cf. Robspierre, Lenin, Stalin, to name but a few leaders)
-- because they need not refer to any type of immutable soul, and no ultimate responsibility.

Also logical `proofs' of the existence of God, from  Anselm of Canterbury onwards to G\"odel and beyond~\cite{benzm-Paleo-2014}
are based on assumption which make them {\em means relative} with respect to these assumptions
and the type of logic used.

What can we learn from this? I suggest to keep in mind that all narratives -- as useful as they may appear in tool-building and technology --
represent no absolute truth but are of preliminary nature, and are subject to constant changes, as time passes by  and history shows.
Presently physics claims that the world consists almost entirely of a vacuum void  which fluctuates by creating and annihilating all kinds of
quanta -- the quanta themselves are point particles with no spatial extension at all; some of them mediate various forces,
some of them carry various charges and capacities to interact.
According to this understanding, a `solid' table is a very special emptiness containing these quanta.
Gravity has been translated into the geometry of space-time.
Many observations in deep space appear unexplained; the metaphorical hypothesis for these puzzling phenomena being {\em dark matter}
(of unknown, probably yet undiscovered, type),
which appears to be invisible, yet makes up most of the universe.
These fragmentary hypotheses and conjectures are hardly a solid basis for a scientific ontology; less a complete, temporally stable body of knowledge!

In coping with Jaynes' ``Mind Projection Fallacy'' \cite{jaynes-89,jaynes-90}
I would personally recommend to adapt a contemplative strategy of {\em evenly-suspended attention}
outlined by  Freud~\cite{Freud-1912}.
It is a manner in which the individual should listen to the universe; without any too strong (mostly unconscious) emotional bias.
(For instance, fear creates tendencies to accept the opposite: fear of determinism yields longing for indeterminism, and {\it vice versa}~\cite{2002-cross}.)
We need to be open for new approaches, scenarios and phenomena; as well as of being aware,
in the spirit of Socrates,
of the vastness of the domains of physical existence we know very little about.
Augustinus' {\it dictum} {\it ``Ei mihi, qui nescio saltem quid nesciam!''}
{ (Alas for me, that I do not at least know the extent of my own ignorance!)}
guides us more than ever.

So, by having this in mind, we are finally in the position to cautiously, and, with hopefully evenly-suspended attention,
engage questions regarding a `room for divinity' in the sciences,
or, conversely, a `room for science' in religion.

\subsection{A rise of indeterminism}

Almost unnoticed at first,  the tide of indeterminism started to build
toward the end of the nineteenth century~\cite{purrington,Kragh-qg}.
At that time, the prevalent mechanistic theories faced an increasing number of anomalies:
to name but a few, there was
Poincar\'e's discovery of  instabilities  of trajectories of celestial bodies
(which made them extremely sensible to initial conditions),
radioactivity~\cite{Kragh-1997AHESradioact,Kragh-2009_RePoss5},
X-rays,
specific heats of gases and solids,
emission and absorption of light, in particular, blackbody radiation,
and
the irreversibility dilemma of statistical physics based on reversible mechanics and electrodynamics.

{\it Fin de si\'ecle} 1900 followed a short period of revolutionary new physics, in particular,
quantum theory and relativity theory,
without any strong metaphysical preference toward either determinism or indeterminism.
Then indeterminism erupted boldly with Born's claim that quantum mechanics has it both ways:
the quantum state evolves strictly deterministically,
whereas the individual event or measurement outcome occurs indeterministically.
Born made it clear that he was\cite[p.~866]{born-26-1}  {\em ``inclined to give up determinism in the world of atoms;''}
that there is no cause for certain individual quantum events;
that is, such outcomes occur irreducibly at random.

Another indeterministic feature of quantum mechanics is {\em complementarity:} there exist collections of observables (such as position and momentum)
which cannot be simultaneously operationalized (i.e. prepared and measured) with arbitrary precision.
Still another indeterministic quantum feature is the {\em value indefiniteness} of at least all but one complementary observables~\cite{specker-60,PhysRevA.89.032109}.

There followed a fierce controversy, with many researchers such as Born, Bohr, Heisenberg, and Pauli
taking the indeterministic stance,
whereas others,
like Planck~\cite{born-55}, Einstein~\cite{epr,ein-reply}, Schr\"odinger, and De Brogli, leaning toward determinism.
This latter position was pointedly put forward by Einstein's {\it dictum} in a letter to Born,
dated December~12, 1926~\cite[113]{born-69}:
{``In any case I am convinced that he [the Old One] does not throw dice.''}

At present, indeterminism is preached by the orthodoxy
to the extend that it is declared {\em ``the message of the quantum''}~\cite{zeil-05_nature_ofQuantum}.
Although such claims are provable unprovable,
they are often motivated by the success of the quantum postulates, as well as
from formal theorems about predictions of general deterministic theories
(relative to some supposedly reasonable assumptions such as omniexistence and contextuality~\cite{svozil-2013-omelette} as well as locality) --
such as Bell's theorem~\cite{Pit-94} and the Kochen-Specker theorem~\cite{specker-60,pitowsky:218,cabello:210401,PhysRevA.89.032109}.

The last quarter of the twentieth century saw the rise of yet another form of physical indeterminism
-- or, rather, unpredictability --
originating in Maxwell's and Poincar\'e's aforementioned
discovery of instabilities of the motion of classical bodies
against variations of initial conditions~\cite{Campbell-1882,poincare14,Diacu96-ce}.

In parallel, G\"odel's incompleteness theorems~\cite{godel1,tarski:32,davis-58,davis,smullyan-92},
as well as related findings in the computer sciences~\cite{turing-36,chaitin3,calude:02,gruenwald-vitanyi},
put an end to Hilbert's program of finding a finite axiom system for all of mathematics.
G\"odel's incompleteness theorems also established formal
bounds on provability, predictability, and induction.
(The incompleteness theorems also put an end to philosophical contentions
expressed by~\cite[101]{schlick-35} that, beyond epistemic unknowables and
the ``essential incompetence  of human knowledge,'' there is ``not a single real
question for which it would be {\em logically} impossible to find a solution.'')

Alas, just like determinism, physical indeterminism cannot be proved, nor can there be given any reasonable criterion for its falsification.
After all, how can one check against all laws and find none applicable?
Unless one is willing to denote any system whose laws are currently unknown
or whose behavior is hard to predict with present techniques as indeterministic,
there is no scientific substance to such absolute claims,
especially  if one takes into account the bounds imposed by the theory of computability~\cite{odi:89,odi:99}.
So both positions --  determinism as well as indeterminism -- must be considered conjectural.

\section{Options in view of {\it Scylla} and {\it Charybdis}}

Like {\it Odysseus} trapped between {\it Scylla} and {\it Charybdis},
our physical worldview, as well as providence and free will, appears to be severely restricted by
physical determinism as well as complete indeterminism.
Does a clockwork universe, as well as one pushing uncontrollable chance,
leave any room for any freedom of the individual, or for divine interaction?

\subsection{Determinism}
Determinism blocks free will by the principle of sufficient reason.
Determinism might be beautiful and
``rich'' in the sense of allowing ornamentation,
but it lacks any kind of {\em steering mechanism}, or {\em freedom of choice.}

Formally, one of the most extreme forms of  determinism is expressed by the unitary quantum mechanical state evolution,
amounting to mere permutations, that is, one-to-one transitions, among states and orthonormal bases~\cite{Schwinger.60}.

\subsection{Emergence of structure from primordial chaos}

{\it Creatio continua,} that is, the {\it ex nihilo} occurrence of events without any cause,
leaves no room for choice either: because if events happen {\it ex nihilo} and uncontrollably,
there is no freedom of choice between alternatives either.

Let us, for the sake of exposing an extreme position, contemplate on an infinite universe without any law.
That is, its behavior cannot be ``compressed'' by any algorithm or rule.
One model of such a universe would be random real.
We assume that the algorithmic incompressibility of encoded microphysical structures
might be a quite appropriate formalization of primordial chaos.
Indeed, while in monotheistic religions usually the natural laws are considered to be created by god,
in Greek mythology and cosmology, \textgreek{q'aoc}, that is,
primordial chaos, has been identified with a deity itself.

How could we, in more formal terms,
imagine the apparent lawfulness of the universe;
despite or ``above'' primordial chaos; and emerging from it?
Maybe Ramsey theory could give us a clue --
because, from a combinatorial point of view,
no matter how irregular some behavior or pattern may be appear, the emergence of
``law and order'' is inescapable.
Indeed, once one is dealing with {\em something} (rather than nothing~\cite{holt-existence}), and
no matter how hard one tries to somehow ``construct'' or create a sequence without correlations,
such correlations are unavoidable~\cite{GS-90}.

So maybe what we call `causality' is just correlations?
In this line of speculation the natural laws could eventually be derivable from Ramsey theory.

As an example, for the sake of emerging orderly logical structures,
let us generalize Landman and Robertson's informal characterization of Ramsey theory \cite[p.~1]{landman+robertson2004}
and suppose that {\em Ramsey theory is the structure of properties under set partitions.}
One of the questions that immediately come up is {\em partition logics}
with its models in automaton logic~\cite{schaller-95}
as well as generalized urn models~\cite{wright,svozil-2001-eua},
and the connections to quantum mechanics~\cite{svozil-2008-ql,DonSvo01,svozil-2002-statepart-prl}.

\subsection{Dualistic interfaces as path through {\it Scylla} and {\it Charybdis}}

Suppose that transcendent agents,
interact with a(n) (in)deterministic universe via suitable {\em interfaces.}
In what follows we shall refer to the transcendental universe as the beyond.


For the sake of  metaphorical models,
take Eccles' mind-brain model~\cite{eccles:papal},
or consider a virtual reality, and, more particular, {\em a computer game.} In such a gaming universe, various human players are represented
by avatars.
There, the universe is identified with the game world created by an algorithm (essentially, some computer program),
and the transcendental agent is identified with the human gamer.
The interface consists of any kind of device and method connecting a human gamer with the avatar.
Like the god {\em Janus} in the Roman mythology, an interface possesses two faces or handles: one into the universe, and a second one into the beyond.

Human players constantly input or inject choices through the interface, and {\it vice versa.}
In this {\em hierarchical, dualistic} scenario, such choices need not solely (or even entirely) be determined
by any conditions of the game world:
human players are transcendental with respect to the context of the game world,
and are subject to their own universe they live in (including the interface).
Alas the game world itself is totally deterministic in a very specific way:
it allows the player's input from beyond; but other than that it is created by a computation.
One may think of a player as a specific sort of indeterministic (with respect to intrinsic means)
{\em oracle}, or subprogram, or functional library.

Another algorithmic metaphor is an {\em operating system},
or a {\em real-time computer system}, serving as context.
(This is different from a classical Turing machine, whose emphasis is not on interaction with some user-agent.
The user is identified with the agent.
Any user not embedded within the context is thus transcendent with respect to this computation context.
In all these cases the  real-time computer system acts deterministically on any input received from the agent.
It observes and obeys commands of the agent handed over to it {\em via} some interface.
An interface could be anything allowing communication between the real-time computer system and the (human) agent;
say a touch screen, a typewriter(/display), or any brain-computer interface.

\section{How to acknowledge intentionality?}

The mere existence of gaps in the causal fabric are no sufficient condition for the existence of providence or free will,
because these gaps may be completely supplied by {\em creatio continua,}

As has already been observed by Frank~\cite[Kapitel~{III}, Sects.~14,~15]{frank},
in order for any {\em miracle} or free will to manifest itself
through any such gap in the natural laws, it needs to be {\em systematic,}
{\em according to a plan}
and
{\em intentional} (German {\it planm\"a\ss ig}).
Because if there were no possibilities to inject information or other matter or content
into the universe from beyond, there would be no possibility to manipulate the universe,
and therefore no substantial choice.

Alas, intentionality may turn out to be difficult or even impossible to prove.
How can one intrinsically decide between chance on the one hand, and providence, or some agent executing free will through the gap interface, on the other hand?
The interface must in both cases employ gaps in the intrinsic laws of the universe,
thereby allowing steering and communicating with it in a feasible, consistent manner.
That excludes any kind of immanent predictability of the signals emanating from it.
(Otherwise, the behavior across the interface would be predictable and deterministic.)
Hence, for an embedded observer~\cite{toffoli:79} employing intrinsic  means
which are operationally available in his universe~\cite{svozil-94},
no definite criterion can exist to either prove or falsify claims regarding mere
chance (by {\it creatio continua}) {\it versus} the free choice of an agent.
Both cases
--
free will of some agent as well as complete chance
--
express themselves by irreducible intrinsic indeterminism.

For the sake of an example, suppose for a moment that
we would possess a sort of {\em `Ark of the Covenant,'}
an oracle which communicates to us the will of the beyond, and, in particular, of divinity.
How could we be sure of that?  (Sarfatti, in order to investigate the paranormal, attempted to built what he called an  {\em  Eccles telegraph} by connecting a
radioactive source to a typewriter.)
This situation is not dissimilar to problems in recognizing hypercomputation, that is, computational capacities beyond universal computation~\cite{2007-hc}.

\section{Physical gaps}

If we translate the algorithmic metaphors mentioned earlier into the context of our own universe,
we have to observe whether all the respective components are physically feasible.
In particular, we might ask the following questions:
(i) Do there exist potential gap-type interfaces in our universe allowing communication with some (supposedly transcendental) agent?
(ii) Are there constraints on such interventions~\cite{maryland,greenberger-svozil,svozil-07-physical_unknowables}?
We may also speculate about the transcendental nature of any agent communicating with our universe {\em via} such interface.

The first question, in particular, the existence of suitable  {\em gaps in the natural laws}
and the causal fabric of the universe, has been investigated by Frank~\cite[Chapter~{III}, Sec.~12]{frank,franke},
as well as by more recent research~\cite{Russel-nioda-1}.
Several physical gap constructions will be critically reviewed next.

\subsection{Deterministic chaos and spontaneous symmetry breaking}

Already in 1873, Maxwell identified a certain kind of {\em instability} at {\em singular points}
as rendering a gap in the natural laws~\cite[211-212]{Campbell-1882}:
{\em ``$\ldots$~when an infinitely small variation in the present state may bring about a finite difference in the state of the
system in a finite time, the condition of the system is said to be unstable.
It is manifest that the existence of unstable conditions renders impossible the prediction of future events, if our
knowledge of the present state is only approximate, and not accurate.''}

\begin{figure}
                \begin{centering}
\unitlength 3mm 
\linethickness{0.4pt}
\ifx\plotpoint\undefined\newsavebox{\plotpoint}\fi 
\begin{picture}(9,7)(0,0)
\thicklines
\put(0,0){\color{blue}\line(1,0){3.0}}
\put(9,0){\color{blue}\line(1,0){3.0}}
\put(3,0){\color{orange}\line(1,2){3.0}}
\put(9,0){\color{orange}\line(-1,2){3.0}}
\put(6,6.9){\color{black}\circle*{2}}
\put(1.5,1){\color{gray}\circle*{2}}
\put(10.5,1){\color{gray}\circle*{2}}
\put(1.5,1){\color{white}\makebox(0,0)[cc]{$0$}}
\put(10.5,1){\color{white}\makebox(0,0)[cc]{$1$}}
\end{picture}
                \end{centering}
                \caption{(Color online) A gap created by a black particle sitting on top of a potential well.
The two final states are indicated by grey circles. Their positions can be coded by $0$ and $1$, respectively.}
                \label{fig:2014-fw-instability}
        \end{figure}
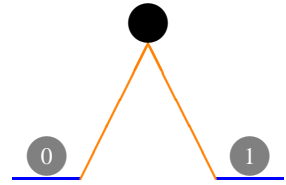

Fig.~\ref{fig:2014-fw-instability} (see also Frank's figure 1 in Chapter~{III}, Section~13) depicts a one dimensional gap configuration envisioned by Maxwell: a
{\em ``rock loosed by frost and balanced on a singular point of the mountain-side, the little spark which
kindles the great forest,~$\ldots$''}
On top, the rock is in perfect balanced symmetry.
A small perturbation or fluctuation causes this symmetry to be broken,
thereby pushing the rock either to the left or to the right hand side of the potential divide.
This dichotomic alternative can be coded by $0$ and by $1$, respectively.

One may object to this scenario of {\em spontaneous symmetry breaking}
by maintaining that, if indeed the symmetry is perfect, there is no movement,
and the particle or rock stays on top of the tip (potential).
Any slightest movement might either result from a microscopic asymmetry or imbalance of the particle,
or from fluctuations of any form, either in the particle's position due top quantum zero point fluctuations,
or by the surrounding environment of the particle.
For instance, any collision of gas molecules with the rock may push the latter over the edge
by thermal fluctuations.

Moreover, {\em deterministic chaos} is not indeterministic at all:
the randomness resides in the {\em assumption} of the continuum from which the initial value is `drawn' (just like an urn).
In this case, almost all (of Lebesgue measure one) initial values
are not representable by any algorithmically compressible number;
that is,  they are random~\cite{MartinLöf1966602,calude:02}.
Deterministic chaos
unfolds the information contained therein by a recursively enumerable (computable),
deterministic evolution function.
If the continuum assumption is dropped, then what remains is Maxwell's
and Poincar{\'e}'s observation of the {\em epistemic} unpredictability
of the behavior of a deterministic system
due to instabilities and diverging evolutions from almost identical initial state.

\subsection{Quantum oracles}

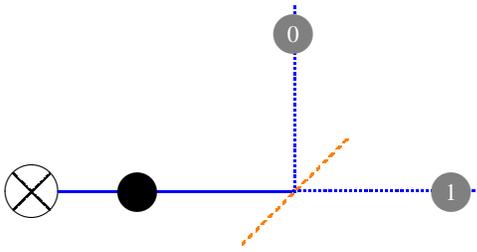
\begin{figure}
                \begin{centering}
\unitlength 0.7mm 
\linethickness{0.4pt}
\ifx\plotpoint\undefined\newsavebox{\plotpoint}\fi 
\begin{picture}(94.75,44.875)(0,0)

\put(10,10){\circle{10}}
\thinlines
\multiput(6.452,6.452)(.033885714,.033671429){210}{\line(1,0){.033885714}}
\multiput(13.523,6.452)(-.033880952,.033671429){210}{\line(-1,0){.033880952}}

\thicklines
{\color{blue}
\put(15,10){\line(1,0){44.5}}
\multiput(59.68,9.93)(.972222,0){37}{{\rule{.8pt}{.8pt}}}
\multiput(59.805,9.805)(0,.972222){37}{{\rule{.8pt}{.8pt}}}
}
\put(30,10){\color{black}\circle*{8}}
\put(59.68,39.93){\color{gray}\circle*{8}}
\put(89.68,9.93){\color{gray}\circle*{8}}
\put(59.68,39.93){\color{white}\makebox(0,0)[cc]{$0$}}
\put(89.68,9.93){\color{white}\makebox(0,0)[cc]{$1$}}

\thicklines
{\color{orange}
\multiput(49.93,-.07)(.0322581,.0322581){20}{\line(1,0){.0322581}}
\multiput(51.22,1.22)(.0322581,.0322581){20}{\line(1,0){.0322581}}
\multiput(52.51,2.51)(.0322581,.0322581){20}{\line(1,0){.0322581}}
\multiput(53.801,3.801)(.0322581,.0322581){20}{\line(0,1){.0322581}}
\multiput(55.091,5.091)(.0322581,.0322581){20}{\line(0,1){.0322581}}
\multiput(56.381,6.381)(.0322581,.0322581){20}{\line(1,0){.0322581}}
\multiput(57.672,7.672)(.0322581,.0322581){20}{\line(1,0){.0322581}}
\multiput(58.962,8.962)(.0322581,.0322581){20}{\line(0,1){.0322581}}
\multiput(60.252,10.252)(.0322581,.0322581){20}{\line(0,1){.0322581}}
\multiput(61.543,11.543)(.0322581,.0322581){20}{\line(0,1){.0322581}}
\multiput(62.833,12.833)(.0322581,.0322581){20}{\line(1,0){.0322581}}
\multiput(64.123,14.123)(.0322581,.0322581){20}{\line(1,0){.0322581}}
\multiput(65.414,15.414)(.0322581,.0322581){20}{\line(0,1){.0322581}}
\multiput(66.704,16.704)(.0322581,.0322581){20}{\line(0,1){.0322581}}
\multiput(67.994,17.994)(.0322581,.0322581){20}{\line(0,1){.0322581}}
\multiput(69.285,19.285)(.0322581,.0322581){20}{\line(0,1){.0322581}}
}
\end{picture}
                \end{centering}
                \caption{(Color online) A gap created by a quantum coin toss. A single quantum (symbolized by a black circle
from a source (left crossed circle)
impinges on a semi-transparent mirror (dashed line), where it is reflected and transmitted with a 50:50 chance.
The two final states are indicated by grey circles. The exit ports of the mirror can be coded by $0$ and $1$, respectively.}
                \label{fig:2014-fw-qcointoss}
        \end{figure}

A quantum mechanical gap can be realized by a {\em half-silvered mirror}~\cite{svozil-qct,stefanov-2000,zeilinger:qct},
with a 50:50 chance of transmission and reflection,
as depicted in Fig.~\ref{fig:2014-fw-qcointoss}.
A gap certified by quantum value indefiniteness necessarily has to operate with more than two exclusive outcomes~\cite{PhysRevA.89.032109}.
Ref.~\cite{2012-incomput-proofsCJ} presents such a qtrit configuration.

One may object to the orthodox view~\cite{zeil-05_nature_ofQuantum} of {\em quantum indeterminism} by pointing out
that it is merely based on a belief without proof.
It is not at all clear where exactly the randomness generated by a half-silvered mirror resides; that is,
where the stochasticity comes from, and what are its origin.
(Often vacuum fluctuations originating from the second, empty, input port are mentioned,
but, pointedly stated~\cite[p.~249]{chau}, these {\em ``mysterious vacuum fluctuations $\ldots$ may be regarded as sugar coating for the bitter pill of
quantum theory.''})

More generally, any irreversible measurement process,
and, in particular,
any associated `collapse,' or, by another denomination, `reduction' of the quantum state (or the wave function) to the post-measurement state
is a postulate which appears to be {\em means relative} in the following sense.

The beam splitter setup is not irreversible at all
because a 50:50 mirror has a quantum mechanical representation as a permutation of the state,
such as a unitary Hadamard transformation;
that is, with regard to the quantum state evolution the beam splitter acts totally deterministic; it can be represented
by a one-to-one function, a permutation.
(Experimentally, this can be demonstrated by serially concatenating two such 50:50 mirrors so that the output ports of the first mirror
are the input ports of the second mirror. The result (modulo an overall phase) is a Mach-Zehnder interferometer reconstructing the original
quantum state.)

Formally -- that is, within quantum theory proper, augmented by the prevalent orthodox `Kopenhagen-type' interpretation --
it is not too difficult to locate the origin of randomness at the beam splitter configuration:
it is (i) the possibility that a quantum state can be in a {\em coherent superposition}
of classically distinct and mutually exclusive (outcome or scattering) states of a single quantum;
and (ii) the possibility that an {\em irreversible measurement} {\it ad hoc} and {\it ex nihilo} stochastically `chooses' or `selects' one of these
classically mutually exclusive properties, associated with a measurement outcome. This, according to the orthodox interpretation
of quantum mechanics, is an irreducible indeterministic many-to-one process --
it transforms the coherent superposition of a multitude of (classically distinct) properties into a single, classical property.
This latter assumption (ii) is sometimes referred to as the {\em reduction postulate.}

Already
Schr\"odinger has expressed his dissatisfaction with both assumptions (i) and (ii), and, in particular, with the quantum mechanical concept of
ontological existence of
{\em coherent superposition}, in various forms at various stages of his life:
he polemicized against (i) by quoting the burlesque situation of a cat which is supposed to be in a coherent superposition between death and life~\cite{schrodinger}.
He also noted the curious fact that, as a consequence of (i) and
in the absence of measurement and state reduction (ii), according to quantum mechanics we all (as well as the physical universe in general),
would become quantum jelly~\cite{schroedinger-interpretation}.

Alas, what in the orthodox scriptures of quantum mechanics often is referred to as `irreversible measurement' remains conceptually unclear,
and is inconsistent with other parts of quantum theory.
Indeed, it is not even clear that, ontologically, an irreversible measurement exists!
Wigner~\cite{wigner:mb} and, in particular, Everett~\cite{everett,everett-collw} put forward ontologic arguments against irreversible measurements
by extending the cut between a quantum object and the classical measurement apparatus to include both object
{\em as well as} the measurement apparatus in a uniform quantum description.
As this latter situation is described by a permutation (i.e. by a unitary transformation),
irreversibility, and what constitutes `measurement' is lost.
Indeed, the reduction postulate (ii) and the uniform unitarity of the quantum evolution cannot both be true, because the former
essentially yields a many-to-one mapping of states, whereas uniform unitarity merely amounts to a one-to-one mapping, that is, a permutation, of states.
In no way can a many-to-one mapping `emerge' from any sort of concatenation of one-to-one mappings!
Stated differently, according to the reduction postulate (ii), information is lost;
whereas, according to the unitary state evolution, no information is ever lost.
So, either one of these postulates must be ontologically wrong (they may be epistemically justified {\em for all practical purposes}~\cite{bell:a1}, though).
In view of this situation, I am (to use Born's dictum~\cite[p.~866]{born-26-1})
inclined to give up the reduction postulate disrupting permutativity, and, in particular,
unitarity, in the world of single quantum phenomena, in favor of the latter; that is, in favor of  permutativity, and, in particular,
unitarity.

The effort to do so may be high, as detailed beam recombination analysis of  a Stern-Gerlach device (the spin analogue of a beam splitter in the Mach-Zehnder interferometer)
shows~\cite{engrt-sg-I,engrt-sg-II}.
Nonetheless, experiments (and proposals) to ``undo'' quantum measurements
abound~\cite{PhysRevD.22.879,PhysRevA.25.2208,greenberger2,Nature351,Zajonc-91,PhysRevA.45.7729,PhysRevLett.73.1223,PhysRevLett.75.3783,hkwz}.
Thus we could say that {\em for all practical purposes}~\cite{bell-a},
that is, {relative to the physical means}~\cite{Myrvold2011237} available to resolve the huge number of degrees of freedom involving a
macroscopic measurement apparatus, measurements {\em appear to be} irreversible, but a close enough look reveals that they are not.
So, irreversibility of quantum measurements appears to be epistemic and means relative, subjective and conventional; but not ontic.
(As already argued by Maxwell, this is just the same for the second law of thermodynamics~\cite{Myrvold2011237}.)

\subsection{Vacuum fluctuations}

As stated by Milonni~\cite[p.~xiii]{milonni-book} and others~\cite{einstein-aether,dirac-aether}, {\em ``$\ldots$~there is no vacuum in the ordinary sense of
tranquil nothingness. There is instead a fluctuating quantum vacuum.''}
One of the observable vacuum effects is the {\em spontaneous emission of radiation}~\cite{Weinberg-search}:
{\em ``$\ldots$~the process of spontaneous emission of radiation is one in which ``particles'' are actually created.
Before the event, it consists of an excited atom, whereas after the event, it consists of an atom in a state of lower energy, plus a photon.''}

Recent experiments achieve single photon production by spontaneous emission~\cite{PhysRevLett.39.691,PhysRevLett.85.290,Buckley-12,Stevenson-spontemi,Sanguinetti},
for instance by electroluminescence.
Indeed, most of the visible light emitted by the sun or other sources of blackbody radiation, including incandescent bulbs,
is due to spontaneous emission~\cite[p.~78]{milonni-book}.

Just as in the beam splitter case discussed earlier the quantum (field theoretic) formalism can be used
to compute (scattering) probabilities
--
that is, expectations for occurrences of individual events,
or mean frequencies for large groups of quanta
--
but remains silent for single outcomes.

Alas, also in the quantum field theoretic case,
unitarity, and thus permutations, govern the state evolution.
Thus, for similar reasons mentioned earlier -- mainly the uniformity of
the validity of unitary quantum evolutions
-- the ontological status of indeterminism remains uncertain.

If we follow the quantum canon, any such emission is an irreducible, genuine instance of
creation coming from nothing {\it (ex nihilo)}; more precisely, in theological terms, the spontaneous emission of light
and other particles amounts to an instance of
{\it creatio continua}.
(This is also true for the stimulated emission of a quantum.)

\begin{figure}
                \begin{centering}
\unitlength 0.6mm 
\linethickness{0.4pt}
\ifx\plotpoint\undefined\newsavebox{\plotpoint}\fi 
\begin{picture}(79.526,40)(0,0)
\thicklines
\put(0,0){\color{blue}\line(1,0){50}}
\put(0,40){\color{orange}\line(1,0){50}}
\put(25,40){\color{gray}\vector(0,-1){39}}
\thinlines
{\color{gray}
\qbezier(30,20)(35,13)(40,20)
\qbezier(70,20)(65,27)(60,20)
\qbezier(50,20)(45,27)(40,20)
\qbezier(50,20)(55,13)(60,20)
\put(70,20){\vector(1,-1){2}}
}
\put(77,20){\color{gray}\circle*{8}}
\put(77,20){\color{white}\makebox(0,0)[cc]{$1$}}
\end{picture}
                \end{centering}
                \caption{(Color online) A gap created by the spontaneous creation of a photon.}
                \label{fig:2014-fw-vacuumfluctuation}
        \end{figure}
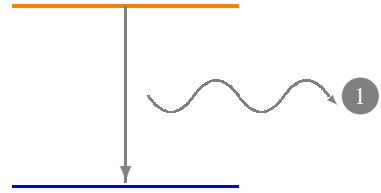

A gap based on vacuum fluctuations is schematically depicted in Fig.~\ref{fig:2014-fw-vacuumfluctuation}.
It consists of an atom in an excited state, which transits into a state of lower energy, thereby producing a photon.
The photon (non-)creation can be coded by the symbols $0$ and $1$, respectively.

It might not be too unreasonable to speculate that all gap scenarios, including spontaneous symmetry breaking and quantum oracles, are ultimately based on vacuum fluctuations.


\section{Interpretations of quantum theory}

When it comes to giving {\it meaning} to the quantum formalism, the majority view seems to be that there is none;
at least none that can be developed rationally:
consider
the ``no interpretation interpretation'' of quantum mechanics suggested by Fuchs and Peres~\cite{fuchs-peres},
or
Feynman's fierce warning~\cite[p.~129]{feynman-law} not to
{\em ``get `down the drain,' into a blind alley from which nobody has yet
escaped.''}

We suggest that any such statement is nothing but a stop sign, or at least a benign recommendation,
erected by frustrated peers suspecting that everybody else will be as unsuccessful as they
have been to figure out how to proceed beyond quantum theory.

At the same time -- and somehow in accord with the no-interpretation interpretation --
the majority also seems to believe in various variants of Bohr's ``Copenhagen interpretation,''
which comes conveniently because Bohr has both been seemingly deep yet not very concrete
-- for the sake of an example take Bohr's
observation~\cite{bohr-1949} regarding
{\em ``the impossibility of any sharp separation between the behavior of atomic
objects and the interaction with the measuring instruments which serve to define the conditions
under which the phenomena appear.''}

Alas Bohr (as so many times before) had it wrong; at least when referring to ``classical measurements:''
because quantum evolution, which is essentially about permutation of states, is {\em inconsistent} with
any irreducible irreversible measurement.
The postulate of any ``fundamental cut'' between observer and object
is, strictly -- but not for all practical purposes (FAPP)~\cite{bell-a} -- speaking,
purely conventional~\cite{svozil-2001-convention}, and contradicts the unitary quantum evolution.
So either quantum theory is universally valid and the Copenhagen interpretation
(at least any one claiming that irreversible measurements exist) is wrong, or the other way round.
(I am inclined to favor quantum theory over the Copenhagen interpretation.)

With regard to Everett's many-world interpretation~\cite{everett-collw} the situation is not dissimilar
to Bohr's informal approach: while fully recognizing the deficiency of any absolute cut
between observer and object~\cite{everett}, Everett is not very specific about branching;
and, in particular, suffers from the impossibility to explain any other than a 50:50
branching ratio at measurements without additional assumptions.

There also exist a huge number of necessarily nonlocal~\cite{epr},
contextual~\cite{kochen1}  ``hidden variable theories'' which, FAPP, cannot be experimentally
falsified, and thus cannot present more in their favor than quantum theory; alas for the overhead of additional
``hidden'' parameters~\cite{CDMTCS458}. (So why bother? Maybe because someday there will be
one with a phenomenology beyond quantum theory.)

Thus, having ruled out, or at least discredited, most so-called ``interpretations''
of the quantum formalism,
at least to this author,
the most natural and compact interpretation of the quantum formalism seems to be that
a pure quantum state -- in the form of a context (that is, a maximal set of co-measurable observables),
and an overlaid two-value frame function~\cite{Gleason} --
is a complete formal representation of a quantized system~\cite{svozil-2013-omelette}.
Through quantum evolution pure states are permuted.
If there is a discrepancy between a state prepared and a property measured;
that is, if there is a mismatch between preparation and measurement,
context translation~\cite{svozil-2003-garda}
introduces epistemic stochasticity
due to the huge number of classical degrees of freedom of the measurement apparatus.

\section{Reprogramming the universe}

So far, all that physics has attempted is preparing physical states and devices capable to manipulate such states in certain ways so that,
by causality, a desirable physical state evolution follows.
In algorithmic terms, this is like feeding the appropriate input into some pre-defined computer,
and processing this input by a pre-defined algorithm to obtain some desired output.
Pointedly stated, so far physics has employed merely a pocket calculator, initially provided -- through {\it creatio ex nihilo} -- by divinity.

A next step would be to {\em change the laws of nature} themselves; that is, in algorithmic language, by {\em reprogramming the universe.}
I suggest to call this type of manipulation {\em ontologic magic} (in contrast to the {\em epistemic magic} performed by professional magicians),
or just {\em magic}.
Of course, magic requires the universe to be programmable; and the natural laws to be mutable.
This, I speculate, might achieve an explanation for the Resurrection of Jesus in modern terms, which, as has been pointed out by Russel,
should not be related to NIODA~\cite[Sect.~3.C.7]{Russel-nioda-1}.

\section{Summary}

We have presented a brief historical account of indeterminism and randomness in physics,
followed by a discussion of the origin and physical sources of physical random number generators
and the associated gaps in the laws of physics.

We have also discussed some philosophical and theological ramifications of {\em creatio continua}; that is,
the creation of randomness {\em ex nihilo} in physics.
{\em Ex nihilo} bits and pieces contradict the principle of sufficient reason.
They also imply gaps in the lawful performance of physical systems.

We have also proposed a (dualistic) ``steering'' mechanism, by which an external agent could interact
through such gaps in the physical laws within an otherwise lawful universe.

In greater detail, the present situation in physics is this:
on the one hand,
virtually all -- and there is not a single exception from this rule --
individual events such as the creation, the scattering, or the annihilation of a quantum (particle),
occur spontaneously, {\it ex nihilo}, and indeterministically.
That means that the details of such events cannot be predicted with arbitrary accuracy by any physical law.
This indeterminism is postulated to be irreducible; that is, it is believed to be ontic rather than epistemic.

On the other hand, if one considers groups of individual events or quanta (particles), or the maximal (that is, pure quantum)
state of knowledge thereof, then the laws of microphysics appear strictly Laplacian deterministically.
There is no room for chance or randomness in the quantum state evolution:
every previous state is a permutation (indeed, a unitary transformation)
from a previous state into a future state.
Stated differently, if quantum mechanics is universally valid, then every
-- and, again, there is not a single exception from this rule -- such
evolution is strictly one-to-one, and onto.

How it is possible to consolidate the quantum (state) reduction (in the
context of quantum wave functions also collapse) postulate as a model for (irreversible) measurements
in an environment which is uniformly governed by pertutations -- that is, unitary evolution -- of the quantum state,
remains unknown.
Indeed,  the maintenance of both the uniform, universal unitary state evolution
on the one hand,
as well as the reduction postulate during irreversible measurements on the other hand,
is inconsistent~\cite{svozil-2013-omelette}.

When it comes to actually locating gaps in the laws of physics, we have discussed three scenarios:
(i) instability of some deterministic evolution due to an extreme sensitivity against variation of initial values
(nowadays often subsumed under the term deterministic chaos);
(ii) quantum beam splitters and state reduction;
(iii) quantum vacuum fluctuations.

In the first scenario one could locate the gap at the assumption
that the initial value is a random real; and that it belongs to a mathematical continuum.
Whether or not this is a viable formalization remains questionable,
in particular, because so far no capacity of the continuum -- such as hyper-Turing computability --
has manifested itself as a physical capacity. The Church-Turing thesis holds as strong as ever.

The second scenario also gives rise to various issues. For instance, it is unclear exactly where the
generation, or rather emergence, of this type of randomness is located.
Is it at the beam splitter? Ideal quantum beam splitters are modeled by Hadamard matrices (up to unitary equivalence),
which are unitary and thus perfectly reversible. Or is it at the detector after the beam splitter?
Or is it ultimately at Wigner's friend~\cite{wigner:mb}?

Actually, I do not believe that such a gap exists ontologically;
the randomness we allegedly observe is means relative to our capacities to resolve the huge number of
degrees of freedom of a ``macroscopic'' measurement apparatus; and hence these gaps in the laws of physics,
as well as physical indeterminism, are epistemic.

The third, field theoretic, scenario also remains problematic, as quantum field theory is just a
quantum theory of many particles; and so inherits many issues within quantum theory.

In any case, in general
it is impossible to either empirically or formally prove or disprove any claims or hypothesis related to physical
determinism or indeterminism;
on the contrary: by reduction to the halting problem (or, equivalently, to Gödel type incompleteness)
it is provable that any claim of either determinism or indeterminism is not provable.
The best one could say is that such claims are means relative: with respect to our current capacities we
are inclined to believe in such-and-such.
Therefore, any absolute statement favoring determinism or indeterminism remains conjectural,  means relative,
context-dependent, as well as preliminary.
But that is true for all human scientific knowledge.

\begin{acknowledgments}
This research has been partly supported by FP7-PEOPLE-2010-IRSES-269151-RANPHYS.
Many discussions about related topics with Alastair A. Abbott and Cristian S. Calude are gratefully
acknowledged.
The manuscript has been prepared for a research grant {\em  SATURN: Scientific and Theological Understandings of Randomness in Nature,}
offered through  Jim Bradley at Calvin College and  Bob Russell at the Graduate Theological Union and  the Center for Theology and the Natural Sciences, Berkeley, California.
All misconception and errors are mine -- {\it Confiteor $\ldots$ quia peccavi nimis cogitatione, verbo, opere et omissione:
mea culpa, mea culpa, mea maxima culpa $\ldots$}
\end{acknowledgments}


%

\end{document}